\documentclass[12pt]{article}
\usepackage[left = 0.8in, right = 0.8in, top = 0.8in, bottom = 0.7in]{geometry}
\usepackage{amsmath}
\usepackage{amssymb}
\usepackage{amsthm}
\usepackage[utf8]{inputenc}
\usepackage{natbib}
\usepackage[colorlinks=true,breaklinks=true,bookmarks=true,urlcolor=blue,
citecolor=blue,linkcolor=blue,bookmarksopen=false,draft=false]{hyperref}
\usepackage{setspace}

\usepackage{tikz}
\usetikzlibrary{decorations.pathmorphing}
\usetikzlibrary{arrows.meta}

\theoremstyle{definition}

\newcommand \re{\mathbb{R}}

\newtheorem{thrm}{Theorem}

\newtheorem{defn}{Definition}

\begin{document}
	
\title{A Simple Characterization of Supply Correspondences\thanks{\protect \emph{Krishnamoorthy:} Carnegie Mellon University, Pittsburgh, USA; \mbox{vinodkri@andrew.cmu.edu}; \emph{Kushnir:} Carnegie Mellon University, Tepper School of Business, Pittsburgh, USA; \mbox{akushnir@andrew.cmu.edu}. We are very thankful to Paul Milgrom for sharing lecture notes on producer choice theory by \cite{LevinMilgromSegal2016}. They inspired this project.}}

\author{Vinod Krishnamoorthy \and Alexey Kushnir}
\date{May 2022}
\maketitle

\begin{abstract}
We prove that supply correspondences are characterized by two properties: the law of supply and being homogeneous of degree zero.\\
	
\noindent \textit{JEL classification:} D21, D24\\[1mm]
\noindent \textit{Keywords: }supply correspondence, supply function, rationalizability, the law of supply, monotonicity, cyclic-monotonicity, homogeneous of degree zero.
\end{abstract}

\section{Introduction}
\label{sec:intro}

We prove that supply correspondences are characterized by two simple properties: \textit{the law of supply} and being \textit{homogeneous of degree zero}. These two properties are both basic properties of profit maximization behavior: The first one captures the intuition that the supply decisions should change in the same direction as prices change. The second property states that if all prices of inputs and outputs are proportionally changed, then the firm supply decision should not change \citep[see Section 5.C in][]{mas1995microeconomic}.

To address the question of when a given supply correspondence is consistent with profit maximization behavior, we consider two concepts of rationalizability. A supply correspondence is \textit{rationalizable} if there exists a convex closed production set such that for every price vector each supply decision is profit maximizing. A supply correspondence is \textit{strongly rationalizable} if it consists of all possible maximizers. Our first result shows that if a supply correspondence satisfies the law of supply and is homogeneous of degree zero, then it is rationalizable. To obtain the strong form of rationalizability, we strengthen the law of supply property. The law of supply is equivalent to a supply correspondence being monotone. Consider the set of all monotone correspondences together with their graphs. A \textit{maximal monotone} correspondence is a monotone correspondence with the graph that cannot be a strict subset of the graph of any other monotone correspondence; that is, its graph is maximal by set inclusion. We show that for a supply correspondence to be strongly rationalizable, it is necessary and sufficient for it to be maximal monotone and homogeneous of degree zero.

The previous literature has provided several characterizations of rationalizable supply functions.\footnote{See \cite{samuelson1948foundations}, \cite{hanoch1972testing}, \cite{Varian1984} for earlier important contributions to production analysis.} Most of these characterizations make strong assumptions about the differentiability of supply function. For example, Proposition 8 in \cite{LevinMilgromSegal2016} states that a continuously differentiable supply function $y$ is rationalizable if and only if it satisfies $Dy(p)p = 0$ and its Jacobian $Dy(p)$ is symmetric, positive semidefinite.\footnote{For a similar statement, see Proposition 7.9 in \cite{kreps2013microeconomic}. \cite{jehle2011advanced} in Chapter 3 reference this result as integrability theorem supply functions, which is parallel to a similar result in consumer choice theory. See also \cite{mas1995microeconomic} and \cite{varian92}.} The condition $Dy(p)p = 0$ is nothing but Euler's law applied to a homogeneous of degree zero supply function. The condition on Jacobian is an analog of the cyclic-monotonicity condition. Our characterization applies to supply correspondences. When supply correspondence satisfies the law of supply, we show that the cyclic-monotonicity condition could be dropped or reduced to the maximal monotonicity condition, depending on what rationalizability concept one is interested in.\footnote{An important result in convex analysis is that if a correspondence is maximal monotone and cyclically monotone, then it is the subdifferential of some proper convex lower semicontinuous function (i.e., the profit function) \cite[see][]{rockafellar2015convex,phelps1997lectures}. In relation to this result, we show that the cyclic-monotonicity condition can be dropped if the function is homogeneous of degree zero. Homogeneous of degree zero is a strong condition, but it is very natural in the context of production theory.} Both conditions are easy to verify. Also, our characterizations does not require any assumptions of differentiablity on the supply function. 

Our characterization of supply correspondences is also closely related to the characterization of combinatorial demand correspondences in the recent paper by \cite{chambers2018characterization}. However, we note several important differences. First, our characterization does not require the image of supply correspondence to be finite (combinatorial). Second, we have different properties: we characterize supply correspondences in terms of the law of supply and homogeneity of degree zero, whereas \cite{chambers2018characterization} characterize combinatorial demand correspondences in terms of the law of demand and upper hemicontinuity. While both the law of demand and the law of supply are two versions of the monotonicity condition with opposite signs, the homogeneity of degree zero condition is absent in consumer choice theory, as the maximization objective is not linear. We also illustrate in Section \ref{sec:discussion} why the maximal monotonicity condition cannot be replaced with the upper hemicontinuity condition when the image of supply correspondence is not finite.

\section{Notation}
\label{sec:notation}

Let us consider an economy with $N\geq 2$ commodities. A typical production plan is a vector  $z=(z_1,...,z_N)\in \mathbb{R}^N$, where an output has $z_n > 0$ and an input has $z_n < 0$. The vector of prices is denoted as $p\in \mathbb{R}^N$.\footnote{We could alternatively assume that the set of possible prices is any cone containing the origin. In particular, we can assume that prices cannot be negative; that is, $p\in\re^N_+$ (see the discussion in Section \ref{sec:discussion}).} We say that for $p, p' \in \mathbb{R}^N$, $p \leq p'$  if each coordinate of $p$ is smaller than the corresponding coordinate of $p'$.

A \textit{supply correspondence} is a mapping $y: \mathbb{R}^N \rightrightarrows \mathbb{R}^N$, giving for each price vector $p$ a set of possible production plans. If $y$ is single-valued, we refer to it as a supply function. Let $\textrm{Im}(y) = \{z \in \mathbb{R}^n\, |\, \exists p \in \mathbb{R}^N \text{ s.t. } z \in y(p)\}$. We consider two concepts of rationalizability.\vspace{2mm} 

\defn Supply correspondence $y: \mathbb{R}^N \rightrightarrows \mathbb{R}^N$ is \textit{rationalizable} if there exists a convex and closed production set $Y$ such that for all $p \in \mathbb{R}^N$,  $y(p) \subseteq \{z\in \re^N\,|\,p\cdot z=\sup_{z' \in Y} p\cdot z'\}.$

\vspace{4mm}
\noindent We note that the requirement for production set $Y$ to be convex and closed is innocuous, because if we find some production set $Y$ to satisfy the maximization condition, the closure of its convex hull will rationalize supply correspondence $y$ as well.\vspace{2mm}

\defn Supply correspondence $y: \mathbb{R}^N \rightrightarrows \mathbb{R}^N$ is \textit{strongly rationalizable} if there exists a convex and closed production set $Y$ such that for all $p \in \mathbb{R}^N$,  $y(p) = \{z\in \re^N\,|\,p\cdot z=\sup_{z' \in Y} p\cdot z'\}.$

\vspace{4mm}
\noindent The first concept requires a supply correspondence to be a subset of all optimal production plans. The second concept demands that the supply correspondence cover all optimal production plans. We also consider three relevant properties for supply correspondences.\vspace{2mm}
\defn[\sc law of Supply; Monotone] A correspondence $y: \mathbb{R}^N \rightrightarrows \mathbb{R}^N$ satisfies \textit{the law of supply (is monotone)} if for all $p, p' \in \mathbb{R}^N$, and all $z \in y(p), z' \in y(p')$, we have $(p - p')\cdot (z-z') \geq 0$. 

\vspace{4mm}
\noindent The law of supply is a basic property of the profit maximization behavior corresponding to the intuition that the quantities should change in the same direction as prices change \citep[see][]{mas1995microeconomic}. At the same time, correspondences that satisfy the law of supply are called simply \textit{monotone} correspondences in the mathematical literature  \citep[see][]{phelps1997lectures}. One could clearly see a justification for this in an application to functions in one-dimensional settings. For one-dimensional settings, the law of supply applied to a single-valued function requires it to be non-decreasing. One could consider the set of all possible monotone correspondences defined on $\re^N$. It is possible to provide a partial order on these correspondences using their graphs and define their maximal elements.

\defn[\sc Maximal Monotone] A subset $G$ of $\re^N\times\re^N$ is said to be \textit{monotone} provided $(z-z')\cdot (p-p')\geq 0$ whenever $(z, p), (z', p')\in G$. A correspondence $y:\re^N \rightrightarrows \re^N$ is \textit{monotone} if and only if its graph
$$
G(y) = \{(z, p) : z \in y(p))\}
$$
is \textit{a monotone set}. A monotone set is said to be \textit{maximal monotone} if it is maximal in the family of monotone subsets of $\re^N\times \re^N$, ordered by inclusion. We say that a correspondence $y$ is \textit{maximal monotone} provided its graph is a maximal monotone set.

\vspace{4mm}
\noindent Finally, we state one more standard property.\vspace{2mm}

\defn[\sc Homogeneity of degree 0] Supply correspondence $y$ is \textit{homogenous of degree 0} if and only if for all $p \in \mathbb{R}^X$, $\lambda > 0$, we have $y(\lambda p) = y(p)$. 

\vspace{4mm}
\noindent We use the above properties in the next section to characterize supply correspondences.

\section{Results}
\label{sec:results}

In this section, we present our main results. First, we show that the law of demand and homogeneity of degree zero are sufficient for rationalizability of supply correspondences. Second, we show that the extension of the law of supply to the maximal monotone condition, and to homogeneity of degree zero are necessary and sufficient for strong rationalizability supply correspondences.

\thrm  A correspondence $y: \mathbb{R}^N \rightrightarrows \mathbb{R}^N$ is \textit{rationalizable} if it satisfies \textit{the law of supply} and is \textit{homogeneous of degree zero}.\footnote{Note that a version of this result for supply functions with a finite image was first established in the working paper by \cite{kushlok2019}.}\label{thrm:weak-rationalizability}
\begin{proof}
	We first show that if  $y: \mathbb{R}^N \rightrightarrows \mathbb{R}^N$ satisfies the law of supply and is homogeneous of degree zero, then $p\cdot z$ is constant over all $z \in y(p)$. To establish it, let $z \in y(p)$, $z' \in y(p)$. From homogeneity of degree zero, we conclude that $z' \in y(\frac{1}{2}p)$. Then, the law of supply implies $(p - \frac{1}{2}p)\cdot (z-z') \geq 0$. Hence, $p\cdot z \geq p\cdot z'$. We can repeat the same argument symmetrically to find $p\cdot z' \geq p\cdot z$, so we must have $p\cdot z =  p\cdot z'$. Therefore, $p\cdot z$ is constant over all $z \in y(p)$.

	We now establish that if  $y$ satisfies the law of supply and is homogeneous of degree zero, it must satisfy the weak axiom of profit maximization; that is, for any $p,p'\in \re^N$, $z \in y(p)$ and $z' \in y(p')$, we must have $p\cdot z \geq p\cdot z'$ \citep[see][]{Varian1984}. The law of supply and homogeneity of degree zero imply that $(p-\lambda p')\cdot (z-z') \geq 0$ for any $\lambda > 0$ or 
	\begin{equation}
		p\cdot (z-z') \geq  \lambda p'\cdot (z-z')\text{ for any } \lambda>0.
		\label{eq:1}
	\end{equation} 
	For the sake of contradiction, assume that $p\cdot (z-z') < 0$. Then, we must also have $\lambda p'(z-z') < 0$ for any $\lambda>0$. We arrive at a contradiction, as we can then find a sufficiently small $\lambda$ such that \eqref{eq:1} is violated. Therefore, $p\cdot (z-z') \geq 0$ or $p\cdot z \geq  p\cdot z'$ for any $p,p'\in \re^N$, $z \in y(p)$ and $z' \in y(p')$. 
	
	Set production set $Y$ to be the closure of convex hull of $\textrm{Im}(y)$. Since linear inequalities are preserved under linear combination and under the operation of taking closure, we obtain that for $z \in y(p)$ and $z' \in Y$, $ p\cdot z \geq p\cdot  z'$. Finally, as product $p\cdot z $ is constant over all $z \in y(p)$, we obtain $y(p) \subseteq \{z \in Y\,|\, p\cdot z = \sup_{z' \in Y} p\cdot z' \}$.	 
\end{proof}

\vspace{4mm}
Note that the law of supply is also a necessary condition for rationalizability \citep[see, e.g.,][]{mas1995microeconomic}.\footnote{An alternative statement of Theorem \ref{thrm:weak-rationalizability} is that any homogeneous of degree one correspondence $y: \mathbb{R}^N \rightrightarrows \mathbb{R}^N$ is \textit{rationalizable} if and only if it satisfies the law of supply.} To see this, note that  rationalizability implies that for any $p, p' \in \re^N, z \in y(p), z' \in y(p')$, $p'\cdot z \leq p\cdot z$. Similarly, $p\cdot z' \leq p'\cdot z'$. Adding these two inequalities together yields $p'\cdot z + p\cdot z' \leq p\cdot z + p'\cdot z'$, which is equivalent to the statement of the law of supply. 

Unfortunately, rationalizable supply correspondences might not satisfy the homogeneity of degree zero, as one could pick maximizers in set $\{z \in Y: p\cdot z = \sup_{z'\in Y}{p \cdot z'}\}$. Hence, it might be the case that $y(p)\neq y(\lambda p)$ for $\lambda >0$. Such a situation is not possible for strongly rationalizable supply correspondences.

We now show our main result that the properties of being maximally monotone and homogeneous of degree zero fully characterize strongly rationalizable supply correspondences.\vspace{2mm}

\thrm  A correspondence $y: \mathbb{R}^N \rightrightarrows \mathbb{R}^N$ is \textit{strongly rationalizable} if and only if it is \textit{maximal monotone} and \textit{homogeneous of degree zero}.
\begin{proof}
	To prove sufficiency, consider $y: \mathbb{R}^N \rightrightarrows \mathbb{R}^N$ that is maximal monotone and homogeneous of degree zero. As any maximal monotone correspondence satisfies the law of supply, Theorem \ref{thrm:weak-rationalizability} implies that $y$ is rationalizable by some convex and closed set $Y$; that is,  $y(p) \subseteq \{z\in \re^N\,|\,p\cdot z=\sup_{z' \in Y} p\cdot z'\}.$ We consider production set $Y$ constructed in the proof of Theorem \ref{thrm:weak-rationalizability} as being the closure of the convex hull of $\textrm{Im}(y)$. It remains to establish that $y(p) = \{z\in \re^N\,|\,p\cdot z=\sup_{z' \in Y} p\cdot z'\}$ for any $p\in \re^N$. 
	
	Assume there exists $(p^*,z^*)$ such that $p^*\cdot z^*=\sup_{z' \in Y} p\cdot z'$ and $z^*\notin y(p^*)$. Hence, for any $p'\in \re^N$ and $z'\in y(p')$, we must have both $p^* \cdot (z^*-z')\geq 0$ and $p'\cdot (z'-z^*)\geq 0$. Therefore, $(z^*-z')\cdot (p^*-p')\geq 0$. Hence, we can construct a new correspondence $y'$ that coincides with $y$ for all $p\neq p^*$ and $y'(p^*)=y(p^*)\cup z^*$. Correspondence $y'$ is monotone and its graph strictly includes the graph of $y$. This contradicts to $y$ being maximal monotone.
	
	To prove necessity, we consider some strongly rationalizable supply function, $y: \mathbb{R}^N \rightrightarrows \mathbb{R}^N$. Hence, there exists a convex and closed set $Y\subset \re^N$ such that $y(p) = \{z\in \re^N\,|\,p\cdot z=\sup_{z' \in Y} p\cdot z'\}$ for any $p\in \re^N$. Let us denote profit function
	$$
	\pi(p)=\sup_{z' \in Y} p\cdot z' \text{ for any } p\in \re^N.
	$$
	Therefore, $y$ is the subdifferential of profit function $y=\partial \pi$. Note that $\pi$ is a proper lower semi-continuous convex function as the support function of set $Y$ \citep[see p. 206 in][]{phelps1997lectures}. Hence, $y$ is maximally monotone as its subdifferential \citep[see Theorem 2.15 in][]{phelps1997lectures}. 
	
	Finally, for any scalar $\lambda>0$, we have
	$$
	y(p) = \{z \in Y: p\cdot z = \sup_{z'\in Y}{p \cdot z'} \} = \{z \in Y: \lambda p\cdot z = \sup_{z'\in Y}{ \lambda p \cdot z'} \}= y(\lambda p), 
	$$
	which shows that any strongly rationalizable supply function $y$ is homogeneous of degree zero. 
\end{proof}

\section{Discussion}
\label{sec:discussion}

\vspace{-2mm}
In this section, we provide additional discussion of our results and more closely relate them to the recent contribution by \cite{chambers2018characterization}.

First, we want to state that the assumption that the supply correspondence is defined on the space of all possible prices $\re^N$ is not necessary for any of our arguments. Our results hold if the supply correspondence is defined on any cone of $\re^N$ containing the origin. In particular, the supply correspondence can be defined on the set of non-negative prices; that is, $y:\re^N_+\rightrightarrows\re^N$.

In addition, our proof does not rely on the result establishing the equivalence between monotone and cyclic-monotone functions for convex domains, which was used by \cite{chambers2018characterization} to analyze the rationalizability of demand correspondences.\footnote{See \cite{bikhchandani2006weak}, \cite{saks2005weak}, \cite{kushnir2021monotone} and references therein for an extensive literature providing conditions on convex and non-convex domains when any monotone function is cyclically-monotone.} This allows us to drop an important assumption in this literature that the image of correspondences is finite.\footnote{Some results that do not require the finite image were considered by \cite{muller2007weak} and \cite{carbajal2015implementability,carbajal2017monotonicity}.} Moreover, the previous literature presents numerous examples illustrating that if one relaxes the assumption of the image being finite, then there is no equivalence between the monotone and cyclic monotone conditions \cite[see, e.g.,][]{rockafellar2015convex,phelps1997lectures,archer2014truthful}. All these examples provide functions that are not homogeneous of degree zero.

Finally, we want to relate our characterization of suppply correspondences more closely to the characterization of combinatorial demand correspondences by \cite{chambers2018characterization}. Both characterizations have different properties: we characterize supply correspondences in terms of maximal monotonicity and homogeneity of degree zero, whereas \cite{chambers2018characterization} characterize combinatorial demand correspondences in terms of the law of demand and upper hemicontinuity. The homogeneity of the degree zero condition is absent in consumer choice theory, as the maximization objective is not linear in prices. The maximal monotonicity condition is a stronger form of the law of supply. Both the law of supply and the law of demand are monotonicity conditions, with the opposite signs reflecting that the supply function should increase in output prices and the demand function should decrease in good prices. At the same time, maximal monotonicity is stronger than uppper hemicontinuity, as any maximal monotone correspondence is upper hemicontinuous, but the reverse direction is generally not true \citep[p. 203 in][]{phelps1997lectures}. 

However, we cannot replace maximal monotonicity with a weaker condition of upper semicontinuity in Theorem \ref{thrm:weak-rationalizability}, as we can then no longer guarantee that the supply correspondence can be strongly rationalized with a convex closed set or even just a closed set (even if it satisfies the law of supply). To illustrate, let us consider supply correspondence $y:\re^2\rightrightarrows\re^2$, defined as 
\begin{equation}
y(p)=\left\{\begin{array}{cl}
	(1,0) & \text{ if } p_1\geq p_2\\
	S     & \text{ if } p_1= p_2\\
	(0,1) & \text{ if } p_2\geq p_1\\
\end{array}\right.,
\label{example}
\end{equation}
\begin{figure}[t]
	\begin{center}
		\begin{tikzpicture}[scale=2.5,domain=0:1]
			\draw[->] (0,0) -- (0,1.5);
			\draw[->] (0,0) -- (1.5,0);
			\draw (0,1.5) node[right=0.5mm] {$y_1$};
			\draw (1.5,0) node[above=0.5mm] {$y_2$};
			
			\draw (0,1) node[left=1mm] {$1$};
			\draw (1,0) node[below=1mm] {$1$};
			\draw[black, dashed] (1,0) -- (0,1);
			\draw[line width=0.8mm,red] (0.3,0.7) -- (0.7,0.3);
			\draw (0.5,0.5) node[above=4mm, right=1mm] {$S$};
			
			\fill[red] (0,1) circle(1pt);
			\fill[red] (1,0) circle(1pt);
		\end{tikzpicture}
	\end{center}
	\vspace*{0mm}{\small {\bf Figure 1.} {The graph of supply correspondence \eqref{example} that satisfies the law of supply, is homogeneous of degree zero, and is upper hemicontinuous, but that cannot be strongly rationalized with a convex or a closed set.}}
	\vspace*{0mm}
\end{figure}
where $S$ is a subset of $Z=\{z\,|\,z=\alpha(0,1)+(1-\alpha)(1,0),\, \alpha\in [0,1]\}$ and we require $(0,1),(1,0)\in S$. Set $S$ is depicted in red in Figure 1. It is straightforward to see that $y$ satisfies the law of supply and homogeneous of degree zero. Hence, $y$ is rationalizable. For example, set $Z$ rationalizes $y$. As $(0,1),(1,0)\in S$ supply correspondence is also upper hemicontinuous. If $S$ is a strict subset of $Z$, then $y$ cannot be strongly rationalized with a convex production set, because such a set has to coincide with $Z$. Moreover, if $S$ is not closed, $y$ cannot be strongly rationalized with a closed set, as such a set has to coincide with $S$. Overall, the above example illustrates that we cannot weaken the maximal monotonicity condition with the upper hemicontinuity condition in the settings when the image of supply correspondence is not finite.

\bibliographystyle{econometrica}
\bibliography{supplyfun}

@Article{chambers2018characterization,
  author    = {Chambers, Christopher P. and Echenique, Federico},
  title     = {A characterization of combinatorial demand},
  number    = {1},
  pages     = {222--227},
  volume    = {43},
  journal   = {Mathematics of Operations Research},
  publisher = {INFORMS},
  year      = {2018},
}

@Article{bikhchandani2006weak,
  author    = {Bikhchandani, Sushil and Chatterji, Shurojit and Lavi, Ron and Mu'alem, Ahuva and Nisan, Noam and Sen, Arunava},
  title     = {Weak monotonicity characterizes deterministic dominant-strategy implementation},
  number    = {4},
  pages     = {1109--1132},
  volume    = {74},
  journal   = {Econometrica},
  publisher = {Wiley Online Library},
  year      = {2006},
}

@Misc{LevinMilgromSegal2016,
  author       = {Jonathan Levin and Paul Migrom and Ilya Segal},
  title        = {Lectures on Producer Theory},
  location     = {Stanford University},
  organization = {Stanford University},
  publisher    = {Stanford University},
  year         = {2016},
}

@Book{mas1995microeconomic,
  author    = {Mas-Colell, Andreu and Whinston, Michael Dennis and Green, Jerry R.},
  title     = {Microeconomic theory},
  publisher = {Oxford university press New York},
  volume    = {1},
  year      = {1995},
}

@Article{phelps1997lectures,
  author    = {Phelps, Robert R.},
  title     = {Lectures on maximal monotone operators},
  number    = {3},
  pages     = {193--230},
  volume    = {12},
  journal   = {Extracta Mathematicae},
  publisher = {Departamento de Matem{\'a}ticas},
  year      = {1997},
}

@Book{rockafellar2015convex,
  author    = {Rockafellar, Ralph Tyrell},
  title     = {Convex analysis},
  publisher = {Princeton university press},
  booktitle = {Convex analysis},
  year      = {2015},
}

@Article{carbajal2017monotonicity,
  author  = {Juan Carlos Carbajal and Rudolf M{\"u}ller},
  title   = {Monotonicity and revenue equivalence domains by monotonic transformations in differences},
  pages   = {29-35},
  volume  = {70},
  journal = {Journal of Mathematical Economics},
  year    = {2017},
}

@Article{carbajal2015implementability,
  author  = {Juan Carlos Carbajal and Rudolf M{\"u}ller},
  title   = {Implementability under monotonic transformations in differences},
  pages   = {114-131},
  volume  = {160},
  journal = {Journal of Economic Theory},
  year    = {2015},
}

@Article{muller2007weak,
  author  = {Rudolf M{\"u}ller and Andr{\'e}s Perea and Sascha Wolf},
  title   = {Weak monotonicity and Bayes--Nash incentive compatibility},
  pages   = {344-358},
  volume  = {61},
  journal = {Games and Economic Behavior},
  year    = {2007},
}

@InProceedings{saks2005weak,
  author    = {Michael Saks and Lan Yu},
  booktitle = {Proceedings of 6th ACM Conference on Electronic Commerce},
  date      = {2005},
  title     = {Weak Monotonicity Suffices for Truthfulness on Convex Domains},
  location  = {New York},
  pages     = {286-293},
  publisher = {ACM Press},
  year      = {2005},
}

@Article{archer2014truthful,
  author  = {Aaron Archer and Robert Kleinberg},
  title   = {Truthful germs are contagious: A local-to-global characterization of truthfulness},
  pages   = {340-366},
  volume  = {86},
  journal = {Games and Economic Behavior},
  year    = {2014},
}

@Article{kushnir2021monotone,
  author    = {Alexey Kushnir and Lev Lokutsievskiy},
  title     = {When is a monotone function cyclically monotone?},
  number    = {3},
  pages     = {853--879},
  volume    = {16},
  journal   = {Theoretical Economics},
  publisher = {Wiley Online Library},
  year      = {2021},
}

@Article{Varian1984,
  author  = {Hal Varian},
  title   = {The Nonparametric Approach to Production Analysis},
  pages   = {579-597},
  volume  = {52},
  journal = {Econometrica},
  year    = {1984},
}

@Article{samuelson1948foundations,
  author  = {Samuelson, Paul Anthony},
  title   = {Foundations of economic analysis},
  number  = {1},
  volume  = {13},
  journal = {Science and Society},
  year    = {1948},
}

@Article{hanoch1972testing,
  author    = {Hanoch, Giora and Rothschild, Michael},
  title     = {Testing the assumptions of production theory: a nonparametric approach},
  number    = {2},
  pages     = {256--275},
  volume    = {80},
  journal   = {Journal of Political Economy},
  publisher = {The University of Chicago Press},
  year      = {1972},
}

@Book{jehle2011advanced,
  author    = {Jehle, G.A. and Reny, P.J.},
  title     = {Advanced Microeconomic Theory},
  isbn      = {9780273731917},
  publisher = {Financial Times/Prentice Hall},
  series    = {The Addison-Wesley series in economics},
  url       = {https://books.google.com/books?id=1c2tbwAACAAJ},
  lccn      = {2010036356},
  year      = {2011},
}

@Book{varian92,
  author    = {Varian, Hal R.},
  title     = {Microeconomic Analysis},
  edition   = {Third},
  publisher = {Norton},
  year      = {1992},
}

@Book{kreps2013microeconomic,
  author    = {Kreps, D.M.},
  title     = {Microeconomic Foundations I: Choice and Competitive Markets},
  publisher = {Princeton University Press},
  series    = {Microeconomic Foundations},
  year      = {2013},
}

@TechReport{kushlok2019,
  author      = {Kushnir, Alexey and Lev Lokutsievskiy},
  institution = {Carnegie Mellon University and Steklov Mathematical Institute of Russian Academy of Sciences},
  title       = {On the equivalence of weak- and cyclic-monotonicity},
  note        = {Working Paper},
  year        = {2019},
}
\end{document}